\documentclass[prl,preprint,amsmath,amssymb]{revtex4-1}

\usepackage{graphicx}
\usepackage{amssymb}
\usepackage{amsfonts}
\usepackage{amsmath}
\usepackage{amsbsy}
\usepackage{natbib}
\usepackage{comment}

\usepackage{color}
\usepackage{float}

\usepackage[colorlinks=true, urlcolor=blue, citecolor=blue, linkcolor=blue]{hyperref}

\renewcommand{\vec}[1]{\mbox{\boldmath$\mathrm{#1}$}}
\let\sb=_ \catcode`\_=\active \def_#1{\ensuremath \sb{\rm#1}}

\renewcommand{\vec}[1]{\mbox{\boldmath$\mathrm{#1}$}}
\newcommand{\be}{\begin{equation}}
\newcommand{\ee}{\end{equation}}
\newcommand{\ben}{\begin{eqnarray}}
\newcommand{\een}{\end{eqnarray}}

\newcommand{\dblue}[1]{{\color{black} #1}}

\begin{document}


\title{Electrically  Tunable   Magnonic Bound States in the Continuum}

\author{Xi-guang Wang$^{1}$, Guang-hua Guo$^{1}$, Jamal Berakdar$^{2}$, Hui Jing$^{3*}$}

\address{$^1$ School of Physics, Central South University, Changsha 410083, China \\
         $^2$ Institut f\"ur Physik, Martin-Luther Universit\"at Halle-Wittenberg, 06099 Halle/Saale, Germany \\
         $^3$ Key Laboratory of Low-Dimensional Quantum Structures and Quantum Control of Ministry of Education, Department of Physics and Synergetic Innovation Center for Quantum Effects and Applications, Hunan Normal University, Changsha 410081, China \\
         $^*$ email: jinghui73@gmail.com}

\date{\today}%

\begin{abstract}
	Low energy excitations of a magnetically ordered system are spin waves with magnon being their excitation quanta. Magnons are demonstrated to be  useful for data processing and communication. To achieve  magnon transport across extended distances, it is essential to minimize  magnonic dissipation  which can be accomplished by material engineering to  reduce  intrinsic damping  or by   spin torques that can counteract damping. 
	This study introduces an alternative methodology  to effectively reduce magnon dissipation based on magnonic bound states in the continuum (BIC). We demonstrate the approach for two antiferromagnetically coupled magnonic waveguides, with one waveguide being attached to a current carrying metallic layer. The current acts on the attached waveguide with a spin-orbit torque  effectively amplifying  the magnonic signal. The setup maps on a non-Hermitian system with  coupled loss and more loss, enabling the formation of dissipationless magnon BIC. We investigate the necessary criteria for the formation  of magnon BIC through electric currents. The influences of interlayer coupling constant, anisotropy constants and applied magnetic field on the current-induced magnon BIC are analyzed. The identified effect can be  integrated  in the design of  magnon delay lines, offering opportunities for the enhancement of magnonic devices and circuits.
\end{abstract}

\maketitle
Magnonic devices employ magnons (the collective excitations in magnetically ordered structures) for information processing and transmission. Advantages include the low energy cost, high speed,  easy miniaturization and integration in spintronic elements.\cite{Chumak2015,Chumak,Cornelissen2015,Kruglyak2010,Lenk2011, Pirro2021} Extensive studies have been carried out to understand the transmission properties and the control mechanisms of magnons in magnetic nanostructures, which resulted in the realization of high-performance magnonic devices such as magnonic logic gates, magnon phase shifters, and magnon transistors.\cite{Chumak9706176, Wang2018, 10.1063/5.0019328,YU20211,Wangq2020,Khitun2010,xiguang2018,PhysRevLett.120.097205} Recent developments make use of  non-Hermitian physics for magnonics and coupled magnonic gain and loss, which enables the emergence of non-Hermitian degeneracy or  exceptional points (EPs).\cite{Wangxinc2020,Liu2019,Yu2020,Zhang2017,Yuan2023,Sui2022,PhysRevB.99.054404,PhysRevLett.131.186705,PhysRevApplied.18.024073,PhysRevResearch.2.013031,Wittrock2024} 
 Based on EPs, several  phenomena occur including  non-reciprocal magnon transmission and enhanced sensitivity at higher-order EPs. \cite{Wangxinc2020,Liu2019,Yu2020,Zhang2017,PhysRevB.99.054404,PhysRevLett.131.186705,PhysRevApplied.18.024073,PhysRevResearch.2.013031}

Here, we deal with BIC which are inherent  to non-Hermitian systems.\cite{Hsu2016,KOSHELEV2019836, PhysRevLett.100.183902, PhysRevA.11.446,PhysRevB.100.115303}  BIC emerge when   resonant modes couple in such a way a mode (BIC) emerges that persists without any radiation losses in a continuous radiation spectrum. Thus, BIC is  a lossless eigenmode, with  purely real  eigenvalue. 
 The existence of BIC has been confirmed, e.g. in photonics \cite{PhysRevLett.100.183902,Kang2023,10.1002/adom.202001469}, acoustics \cite{ChenYuan2016,Yu2022,Liu2022}, electronics \cite{PhysRevB.85.115307,ALVAREZ20151062}. Also, in cavity magnonics magnon-photon coupling  enables  BIC.\cite{PhysRevLett.125.147202, doi:10.1126/sciadv.adg4730} The BIC is  potentially useful   in sensing \cite{doi.org/10.1002/adfm.202104652, Jahani2021, WangHanDuQin+2021+1295+1307}, filtering \cite{Doskolovich19, GaoXingwei2019}, and lasing \cite{Kodigala2017, Hwang2021, Yuyi2021}.
 
\begin{figure}[htbp]
	\includegraphics[width=0.8\textwidth]{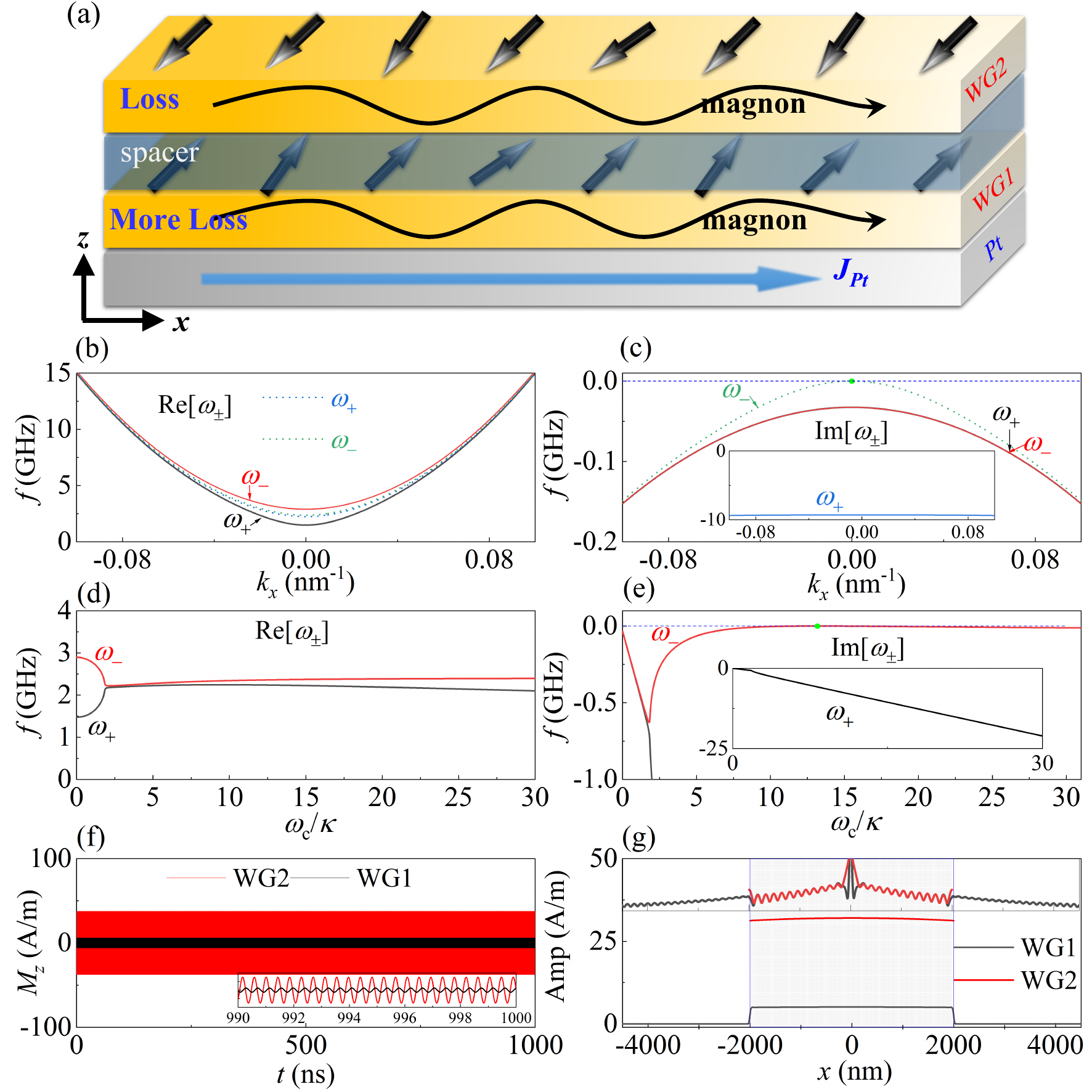}
	\caption{\label{model} (a) Schematic of two antiferromagnetically coupled magnon waveguides on a heavy metal substrate with a large spin Hall angle.  The structure is shown to support BIC. Injecting a charge current $ \vec{J}_{\mathrm{pt}} $ in the metal layer results in a spin orbit torque on the magnetization in the neighboring waveguide (WG1), which effectively enhances  the magnon damping ("more loss"). Magnons in the other waveguide (WG2) experience intrinsic Gilbert damping only ("loss").  Magnons are launched locally at the center ($ x = 0 $) of WG1 or WG2. (b-c) For the charge current $\omega_c = 0$ (solid lines) and $\omega_c = 13.2 \kappa$ (dots), (b) real and (c) imaginary parts of two magnon eigenfrequencies $f = \omega/(2\pi)$ as functions of the wave vector $k_x$. (d) Real and (e) imaginary parts of $f(k_x=0)$ as $\omega_c / \kappa$ varies. The green points signal the BIC. The insets provide enhanced imaginary parts of $\omega_+$ mode. (f) Temporal profile of the $M_z$ component located at the center ($ x = 0 $). The insets are enlarged views of magnetization oscillations. (g) When WG1 (in $ -6000 {\rm nm} < x < 6000 {\rm nm} $) coupled to WG2 exclusively in $ -2000 {\rm nm} < x < 2000 {\rm nm} $, spatial profile of magnon amplitude in two waveguides. Here, the magnon with $f = 2.35$ GHz and $k_x = 0$ is sustained by $\omega_c = 13.2 \kappa$ without external excitation. The inset of (g) is for the case of $\omega_c = 0$.  }
\end{figure}

This paper points out the existence of magnonic BIC in antiferromagnetically coupled magnetic waveguides attached to a metallic layer.
BIC and associated features are  shown to be externally controllable by the current density strength in the metallic layer. We consider modes of two magnonic waveguides that are coupled via  Ruderman-Kittel-Kasuya-Yosida (RKKY) antiferromagnetic interaction (cf. Fig. \ref{model}), a case which is referred  to as a synthetic antiferromagnet (SAF).\cite{syAFM} In comparison to the antiferromagnetic lattice, the SAF demonstrates substantially weaker antiferromagnetic coupling. Yet,  benefits are related to the  structure tunability and easier detectability of magnetic moments which is advantageous in various applications requiring precise magnetic control and detection.\cite{syAFM, Lavrijsen2013}
   The coupled magnonic layers support  two magnonic eigenmodes  homogeneously  damped  due to  'magnon losses'. When the SAF structure is attached to  metallic layer with strong spin orbit coupling, like platinum, due to spin Hall effect, a current flowing in the metallic layer acts with a spin-orbit torque (SOT) on the neighboring magnetic layer (assumed to be insulating here). For a specific direction of the current flow, SOT can enhance the magnon damping in the magnetic layer, i.e. causing  "more magnon loss".\cite{Krivorotov228, Garello2013, Hoffmann2013, Collet2016} Our findings suggest that by manipulating the current density, the system encompassing coupled loss and additional loss can be effectively steered towards the EP and BIC states. We conducted an extensive and detailed analysis on the prerequisites for the appearance and the way of  to steer   BIC and its distinct  features. Exploiting the acquired BIC, we demonstrate how to  construct  magnon delay device  contributing thus to the ongoing  advancement in the design of magnonic logic devices and circuits.

The SAF with two insulating magnetic layers (separated by a thin nonmagnetic spacer) with antiferromagnetic RKKY coupling is depicted in Fig. \ref{model}(a). With appropriate material parameters (such as thickness) for the non-magnetic layer between the waveguides, one can select the case with opposite local magnetization directions $\vec{m}_{1,2}$ in WG1 and WG2. A charge current flowing in the Pt layer (solely attached to WG1) along $x$ direction generates a SOT \dblue{$ \vec{T}_1 = \gamma c_J \vec{m}_1 \times \vec{y} \times \vec{m}_1 $} in WG1, while in WG2 the torque $\vec{T}_2$ amplitude is 0. The SOT strength $ \gamma c_J $ is proportional to charge current density $J_e$ and spin Hall angle $\theta_{SH}$. The magnon dynamics of interest here is well  described by starting from the Landau-Lifshitz-Gilbert (LLG) equation,\cite{Krivorotov228, Collet2016}
\begin{equation}
\displaystyle \frac{\partial \vec{m}_p}{\partial t} = - \gamma \vec{m}_p \times \vec{H}_{\rm{eff,}p}+  \alpha \vec{m}_p \times \frac{\partial \vec{m}_p}{\partial t} + \vec{T}_p.
\label{LLG}
\end{equation}
Here, $p = 1, 2$ corresponds to waveguides WG1 and WG2. $\gamma$ is the gyromagnetic ratio, and $\alpha$ is the intrinsic Gilbert damping. The effective field $ \vec{H}_{\rm{eff},p} = \frac{2 A_{\mathrm{ex}}}{\mu_0 M_s} \nabla^2 \vec{m}_p - \frac{J_{\rm{AF}}}{\mu_0 M_s t_p} \vec{m}_{p'} + \frac{K_y}{\mu_0 M_s} m_{p,y} \vec{y} - \frac{K_z}{\mu_0 M_s} m_{p,z} \vec{z} $  consists of the internal exchange field (exchange constant $ A_{\mathrm{ex}} $), the interlayer antiferromagnetic coupling field (coupling constant $ J_{\rm{AF}} $), easy-axis anisotropy field along $ \vec{y} $ (easy-axis anisotropy constant $ K_y $), in-plane anisotropy field perpendicular to $ x $-$ y $ plane (in-plane anisotropy constant $ K_z $). Here, $ p,p' = 1,2 $ and $ p \ne p' $, $ t_p $ is the $ p $th layer thickness, $M_s$ is the saturation magnetization, and $ \mu_0 $ is the vacuum permeability. 

For numerical calculations and simulations, we use Yttrium iron garnet (YIG) for the WG1/2 with the following magnetic material parameters,   $M_s = 1.4 \times 10^5$ A/m, $A_{ex} =  3 \times 10^{-12}$ J/m, $K_y = 900 $ J/m$^3$, and $K_z = 1.1 \times 10^4$ J/m$^3$. The interlayer antiferromagnetic coupling exchange constant is $J_{AF} = 1.75 \times 10^{-5}$ J/m$^2$, corresponding to $\frac{J_{\rm{AF}}}{\mu_0 M_s t_p} \approx 2.0 \times 10^{4} $ A/m for  $t_p = 5$ nm. The used intrinsic Gilbert damping $\alpha = 0.01$   depends on the  quality of YIG  and can be one or two orders of magnitude smaller. Technical details of numerical simulations are provided in the Supplementary Material (SM).

 \begin{figure}[htbp]
	\includegraphics[width=0.85\textwidth]{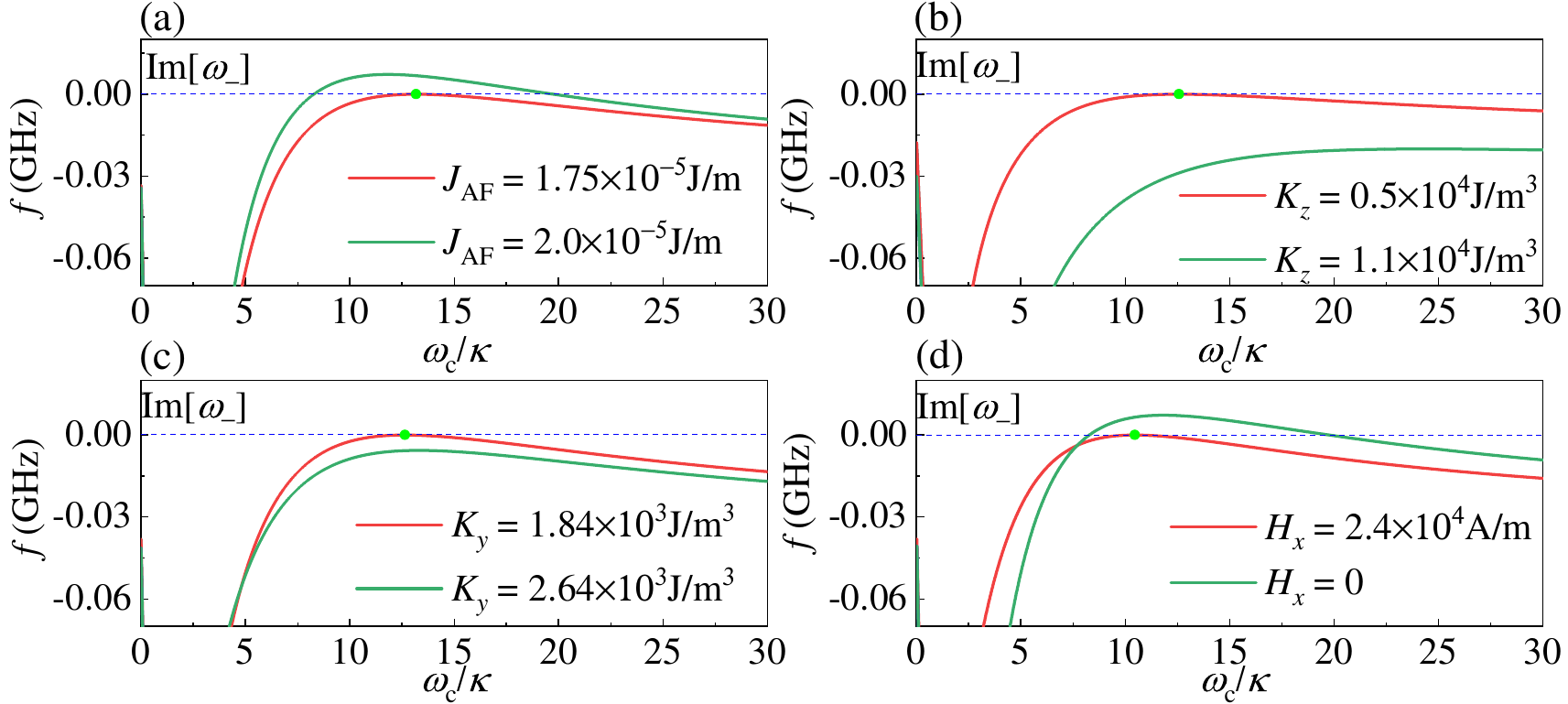}
	\caption{\label{kxbic} For different parameters (a) $ J_{\rm{AF}} $, (b) $K_z$ with $ J_{\rm{AF}} = 0.9 \times 10^{-5} $ J/m, (c) $K_y$ and (d) $H_x$ with $ J_{\rm{AF}} = 2 \times 10^{-5} $ J/m,  the imaginary parts of the $\omega_-$ mode as the WG1 electric current term $\omega_c/\kappa$ varies. Here,  the magnon with $k_x = 0$ is analyzed, and the green points signal the BIC.}
\end{figure}

Magnons are modes of a linearized  version of LLG equations and can be obtained analytically. For this  we consider  small deviations of $\vec{m}_p = (\delta m_{x,p}, 0, \delta m_{z,p}) e^{i(k_x x - \omega t)}$ around the initial antiferromagnetic equilibrium $ \vec{m}_1 = (0, 1, 0) $ and $ \vec{m}_2 = (0, -1, 0) $. Introducing the wave function $\Psi_p^{\pm} =  \delta m_{x,p} \pm i \delta m_{z,p} $, and inserting it into LLG equation (\ref{LLG}), we deduce an eigenvalue equation $\omega \vec{\Psi} = \hat{H} \vec{\Psi}$ with $\vec{\Psi} = (\Psi_1^+, \Psi_1^-, \Psi_2^+, \Psi_2^-)$, and the $4 \times 4$ Hamiltonian $ \hat{H} $ is,
\begin{equation}
\begin{small}
\begin{aligned} 
\displaystyle \hat{H}_0 = \left( \begin{matrix} \omega_0^- - i \omega_J^- & -\omega_z^- & \kappa^- & 0 \\ \omega_z^+ & -\omega_0^+ - i\omega_J^+  & 0 & -\kappa^+ \\-\kappa^+ & 0 & -\omega_0^+ & \omega_z^+ \\ 0 & \kappa^- & -\omega_z^- & \omega_0^- \end{matrix} \right).
\label{ham}
\end{aligned} 
\end{small}
\end{equation}
Here, we use the following notations:  $\omega_z = \frac{\gamma K_z}{(1+\alpha^2)\mu_0 M_s} $,
 $ \kappa = \frac{\gamma J_{\rm{AF}}}{(1+\alpha^2)\mu_0 M_s t_p} $, 
 $\omega_0 = \omega_{ex} + \omega_z + \kappa$, $\omega_{ex} = \frac{\gamma}{1+\alpha^2}(\frac{2A_{ex}k_x^2}{\mu_0 M_s} + \frac{2K_y}{\mu_0 M_s} )$,
$\omega_0^{\pm} = (1 \pm i \alpha) \omega_0$, $\omega_J^{\pm} = (1 \pm i \alpha)\omega_c$, $\omega_c = \frac{\gamma c_J}{1+\alpha^2}$,
 $\omega_z^{\pm} = (1 \pm i \alpha) \omega_z $ and $\kappa^{\pm} = (1 \pm i \alpha) \kappa$.

For the Hamiltonian $ \hat{H} $, we find two frequencies (with positive real parts) corresponding to right-hand precessions around their ground states, while the other two negative frequencies belong to the left-hand precessions. These two positive frequencies are identified as optical and acoustic magnon modes. Without charge current, meaning  $\omega_c = 0$, the magnon dispersion is shown in Figs. \ref{model}(b-c). For all wave vectors $k_x$, one branch of magnon modes always has higher frequencies. With the increase in $k_x$, the gap between two modes is gradually narrowing. The imaginary parts of the two magnon modes are consistently identical for different $k_x$.

While increasing the charge current density in the metallic layer, i.e. increasing  $\omega_c$, we observe the occurrence of EP which is a consequence  of the interplay of magnonic  loss and the additional loss. As demonstrated in Figs. \ref{model}(d-e), the EP (around $ \omega_c = 1.85 \kappa $) is a witness of the convergence of the real parts of two magnon modes (for an enlarged view of EP we refer to SM). Further increasing $\omega_c$, we observe an evident divergence in the imaginary parts of the two modes.  The findings concerning general features of EP are in line with  literature.\cite{Wangxinc2020,Liu2019,Yu2020,Zhang2017,PhysRevB.99.054404,PhysRevLett.131.186705,PhysRevApplied.18.024073,PhysRevResearch.2.013031} Note, the anti-parallel magnetization directions are always stable around the EP, allowing so for the generation of  BIC above EP. This is contrasted with EP-induced instability resulting from coupled gain and loss.\cite{PhysRevB.107.L100402,PhysRevB.108.054428} A detailed discussion for the enhanced loss induced EP is in SM.

In the following, we focus on the magnonic BIC. Beyond the EP threshold, as the current $\omega_c$ induces increased loss in WG1, the divergence in the imaginary parts of the two modes leads to a decrease in the imaginary part of one mode and an increase in the other, as demonstrated by Figs. \ref{model}(d-e). This ultimately results in the imaginary part of one magnon mode $\omega_-$ becoming zero at $ \omega_c = 13.2 \kappa $, indicating BIC. Above this specific BIC, the imaginary component of the magnon mode turns negative again, and no magnon amplification or related instability occurs in the entire range. A similar connection between EP and BIC in non-Hermitian systems was reported in Optics, Cavity Magnonics, et al.\cite{PhysRevLett.125.147202} The magnonic BIC exhibits a dependence on the wave vector $k_x$. As demonstrated by Figs. \ref{model}(b-c), the BIC occurs only at $k_x = 0$ and does not appear for other $k_x$. Choosing a current density at the BIC $ \omega_c = 13.2 \kappa $, the initially injected magnon at $k_x = 0$ can be sustained for a long duration without significant amplification or attenuation, as detailed in Fig. \ref{model}(f). As the imaginary part of magnons near the BIC changes slowly with $\omega_c$, fluctuations in the current density do not impact the experimental observation of BIC-related effects. Furthermore, when the coupling region is restricted  to a specific area, the magnonic BIC is confined inside the coupled region without emission leakage, as demonstrated by the magnon amplitude profile in Fig. \ref{model}(g). When the current is turned off, the injected magnons of the same frequency can effectively spread outside the coupling region (inset of Fig. \ref{model}(g)). Besides, a general feature of a  BIC mode is the  enhanced excitation amplitude.\cite{Hsu2016,PhysRevLett.123.253901, Cai21} In line with this finding, we confirm for our system a significantly enhanced dynamic magnetic susceptibility near the magnonic BIC (see SM).

Also, we analyzed the effects of different magnetic parameters on the magnonic BIC. In Fig. \ref{kxbic}(a), as the coupling constant $ J_{\rm{AF}} $ increases, the imaginary part ${\rm Im}[\omega_-]$ shifts to the positive direction. At $ J_{\rm{AF}} = 2 \times 10^{-5} $ J/m, two dissipationless BIC modes are observed, and the positive magnon imaginary part appears between the two BICs, which can cause magnon amplification and magnetization instability within this range. A more detailed discussion on the two BICs is provided in the SM. Decreasing $ J_{\rm{AF}} $ moves ${\rm Im}[\omega_-]$ towards the negative side, and at $ J_{\rm{AF}} = 0.9 \times 10^{-5} $ J/m the BIC disappears, as proved by the curve in Fig. \ref{kxbic}(b). Reducing the anisotropy constant $K_z$ shifts ${\rm Im}[\omega_-]$ to the positive direction, and a suitable choice of $K_z$ brings the system to a single BIC (Fig. \ref{kxbic}(b)). The anisotropy constant $K_y$ also affects the BIC, as demonstrated by Fig. \ref{kxbic}(c). Moreover, applying a magnetic field $H_x$ along the $x$ axis alters the antiparallel magnetization, leading to a spin-flop state (see SM) and change in the BIC feature. As shown in Fig. \ref{kxbic}(d), with $H_x = 2.4 \times 10^4$ A/m, the two BICs at $ J_{\rm{AF}} = 2 \times 10^{-5} $ J/m merge into a single BIC, avoiding instability of the system. In practice, the value of $ J_{\rm{AF}} $ can be modified by adjusting the thickness of the non-magnetic layer between the waveguides, and the anisotropy constants $K_y$ and $K_z$ can be tuned by adjusting the magnet shape or applying strain. If the system contains multiple BICs and unstable regions, an external magnetic field can be applied to generate a single BIC. These features are helpful in the related experimental observation. 

By requiring  zero dissipation, we find the existence condition of a single BIC, namely 
\begin{equation}
\begin{small}
\begin{aligned} 
\displaystyle &\kappa^4 - 12\alpha^2\omega_0^2\omega_1^2 -  \omega_{2}^2 = 0,\\
\rm{with} \ &\omega_{1}^2 = \omega_0^2-\kappa^2 - \omega_z^2,\\
&\omega_{2}^2 =  [-\omega_3^6 + \frac{1}{2}\sqrt{-4(\kappa^4-12\alpha^2\omega_0^2\omega_1^2)^3+4\omega_3^{12}}]^{1/3},\\
&\omega_{3}^6 =  \kappa^6 - 54 \alpha ^4\omega_0^4(\omega_0^2 - \omega_z^2) + 18 \alpha^2 \kappa^2 \omega^2 (\omega_z^2 - \omega_1^2).
\label{BIC1}
\end{aligned} 
\end{small}
\end{equation}
The electric current value $\omega_c$ of the BIC can be estimated from
\begin{equation}
\begin{small}
\begin{aligned} 
\displaystyle \omega_{BIC} = \frac{\kappa^4 - 12\alpha^2\omega_0^2\omega_1^2 + \kappa^2 \omega_2^2}{6 \alpha \omega_0 \omega_2^2}.
\label{BIC2}
\end{aligned} 
\end{small}
\end{equation}

These results suggest that enhancing loss in WG1 through SOT can reduce effective magnon dissipation, thereby enabling the realization of BIC without loss or even magnon with positive gain. The origin of the magnonic BIC is similar to the Friedrich-Wintgen (FW) BIC, which arises from coupling of different resonances. The FW BIC condition is achieved by tuning the parameters of a non-Hermitian Hamiltonian of two connected resonant structures.\cite{Hsu2016,Yu2022,PhysRevLett.125.147202} Here, two separated resonators  (WG1 and WG2) host modes with different radiation rates, that are coupled via the interlayer coupling, generating a non-Hermitian Hamiltonian. The magnonic BIC condition is achieved by tuning the loss parameter $ \omega_c $ without introducing gain, which causes one lossless BIC mode with the other mode becoming "more lossy". With these features, despite Eq. (\ref{ham}) not being a typical FW Hamiltonian, the magnonic BIC can be regarded as an unconventional BIC. From a different viewpoint, the induced low loss mode can be attributed to effective gain from entanglement between the coupling term ($ J_{\rm{AF}} $) and the off-diagonal imaginary energy in the Hamiltonian Eq. (\ref{ham}). We also note that no similar BIC phenomena were observed in ferromagnetic coupling waveguides, as detailed in the SM. Besides, previous studies have demonstrated that dynamical interference or chiral pumping effects can also sustain the trapped magnons.\cite{PhysRevB.102.054429, Chen2022jpd, Wang2021nano} These effects are different from the magnonic BIC produced by the spin-torque-driven non-Hermitian effect.

\begin{figure}[htbp]
	\includegraphics[width=0.9\textwidth]{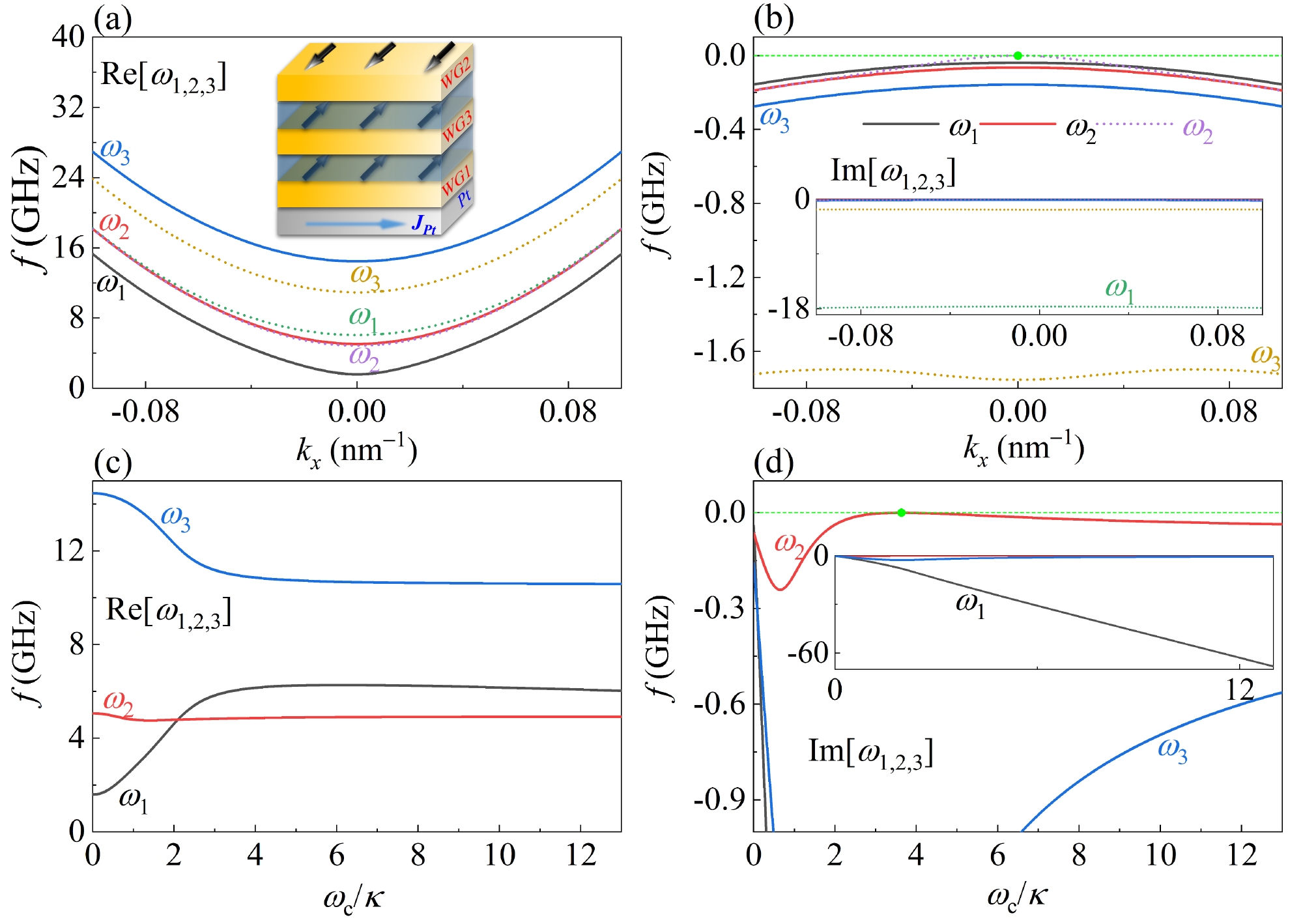}
	\caption{\label{three} Using the three layers model in the inset of (a), for the charge currents $\omega_c = 0$ (solid lines) and $\omega_c = 3.6 \kappa$ (dots), (a) real and (b) imaginary parts of the three magnon eigenfrequencies $f = \omega/(2\pi)$ as functions of the wave vector $k_x$. (c) Real and (d) imaginary parts of $f(k_x=0)$ as $\omega_c / \kappa$ varies. The green points signal the BIC. The insets provide enlarged views of the imaginary parts of $\omega_1$ mode. Here, WG1 (WG3) is coupled to WG2 via a ferromagnetic (antiferromagnetic) coupling $J_{AF} = J_{F} = 1.3 \times 10^{-4}$ J/m$^2$, with $K_y = 900 $ J/m$^3$ and $K_z = 6.2 \times 10^3$ J/m$^3$.}
\end{figure}

To further illustrate the concept of magnonic BIC, we introduce a three layers model, as shown in the inset of Fig. \ref{three}(a). Here, the bottom layer with SOT-enhanced dissipation and the top layer with normal dissipation are both coupled to the middle magnetic layer. Phenomenologically, the bottom and top magnon resonators simultaneously emit magnons into the same channel (the middle layer), resulting in an effective coupling. This scenario closely resembles the FW model reported in Ref. \cite{Hsu2016,Yu2022,PhysRevLett.125.147202}. Following the same analytical method as for the above two layers model, we obtain the results for three layers magnon modes in Fig. \ref{three}. These results confirm that BIC still exists in this system.

\begin{figure}[htbp]
	\includegraphics[width=0.9\textwidth]{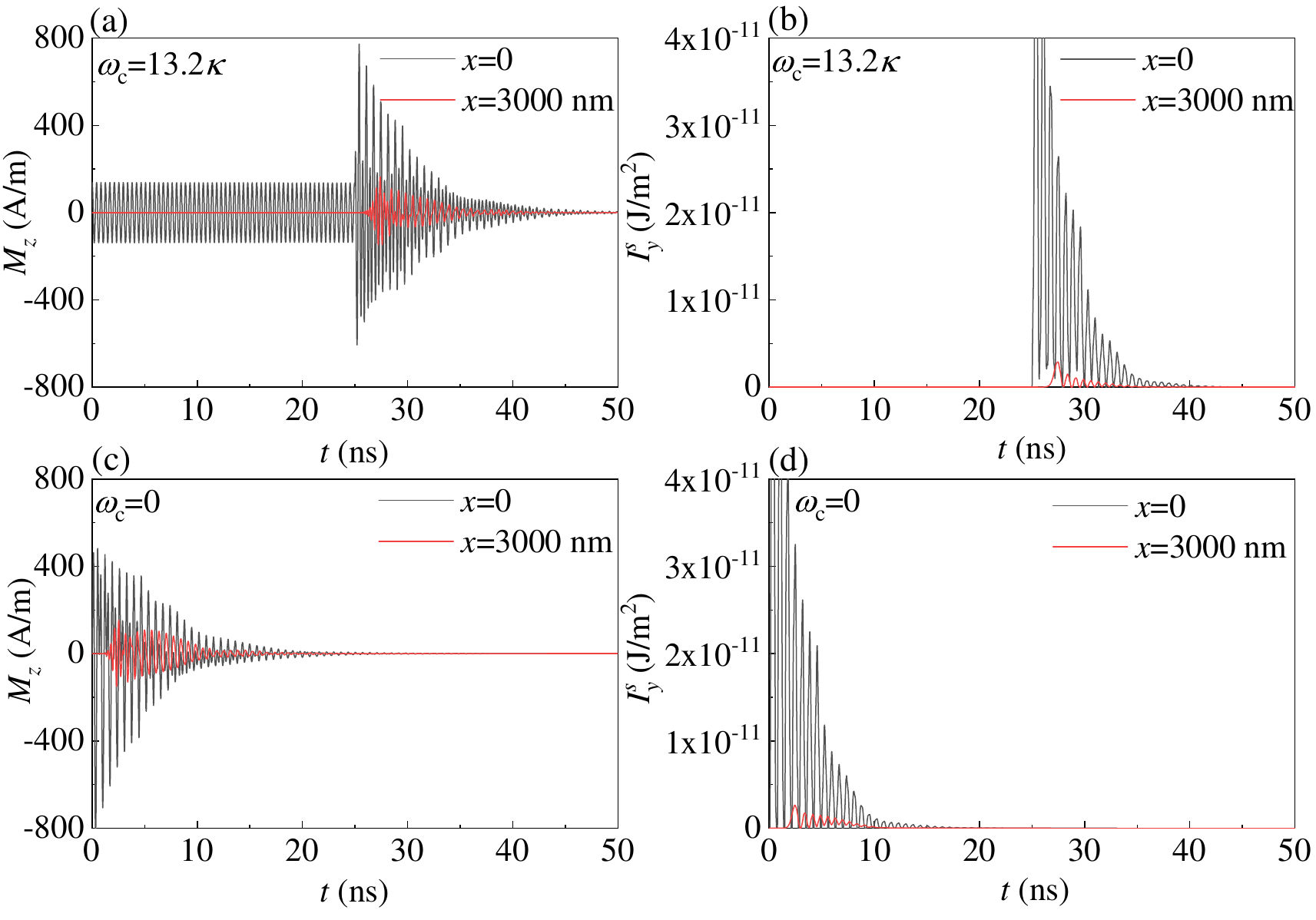}
	\caption{\label{sim} When WG1 (in $ -6000 {\rm nm} < x < 6000 {\rm nm} $ with SOT enhanced magnon loss) is coupled to WG2 exclusively in the region $ -2000 {\rm nm} < x < 2000 {\rm nm} $, the time-resolved oscillations of $M_z$ and spin pumping current $I_y^s$ under (a-b) $\omega_c = 13.2 \kappa$ and (c-d) $\omega_c = 0$ at $x = 0$ and $x = 3000$ nm in WG1.  The electric current with $\omega_c = 13.2 \kappa$ is injected over $ 0 < t < 25 $ ns, and the magnon with $f = 2.35$ GHz is initially excited at $t = 0$.}
\end{figure}

The features of the BIC can be exploited in the design of precise control of magnon signals in delay lines. By adopting a structure where the coupling between WG1 and WG2 is confined to a specific region, the BIC point $\omega_c = 13.2 \kappa$ can effectively trap the non-decaying magnon inside the coupling range, as evidenced by the variations in  Fig. \ref{sim}(a). Thus, magnon transmission toward the right side (e.g., $x = 3000$ nm) is limited. At $t = 25$ ns, the termination of the current leads to the resumption of typical dissipation patterns for magnons, allowing the magnons to enter the right side. For experimental detection, the magnon signal can be detected, for example, by converting it to an electric signal exploiting the spin pumping effect. To this end, we evaluate the spin pumping current via $\vec{I}^s = \frac{\hbar g_r}{4 \pi} (\vec{m} \times \frac{\partial \vec{m}}{\partial t})$. Here, $g_r = 7 \times 10^{18} {\rm m^{-2}}$ is the rescaled interface mixing conductance. As shown in Fig. \ref{sim}(b), the $y$ component of the pumping current $\vec{I}^s$ effectively reflects the variation in the magnon amplitude. The pumping current can be further converted into a detectable electric signal through the inverse spin Hall effect. Compared to the case without electric current $ \omega_c = 0$,  the implementation of the current pulses offers a precise method for engineering the delayed release of magnons. We note that the exact time at which the magnons exit the coupling region is determined by the current pulse; for example, terminating the current pulse at $t = 45$ ns causes the magnons to propagate to the right at that moment (see SM). These findings can be of use in the design of magnonic devices and integrated circuits that feature dynamically modulated losses controlled by electric currents. Compared to the SOT induced dissipationless mode via the anti-damping effect,\cite{Collet2016} the non-Hermitian magnonic BIC here can effectively eliminate magnon damping without triggering instability and nonlinear effects, bringing unique advantages in related applications.

To conclude, we predict  the formation of a magnonic BIC without dissipation in  two antiferromagnetically coupled magnonic waveguides on a current-carrying heavy metal substrate. The engineering of the magnonic BIC is achievable through manipulation of the current density, providing a tunable approach to the control of magnonic states. The existence of BIC is affected by magnetic parameters including the interlayer coupling constant, anisotropy constants, and applied magnetic field. These phenomena can be utilized to design magnonic delay lines that are relevant for the realization of next-generation logic devices and integrated circuits for magnon-based information transmission and processing.

\end{document}